\documentstyle[12pt]{article}
\long\def\del#1\enddel{ }
\let\3=\ss \catcode`\"=\active \let"=\"
  \oddsidemargin=\evensidemargin
\let\ni=\noindent  
 \let\msk=\medskip

\let\a=\alpha

\def\0{\over }    \def\1{\vec }   \def\2{{1\over2}} \def\3{{\ss}}
\def\4{{1\over4}} \def\5{\bar }   \def\6{\partial } \def\7#1{{#1}\llap{/}}
\def\8#1{{\textstyle{#1}}}        \def\9#1{{\bf {#1}}}
\def\_#1{$\underline{\hbox{#1}}$} \def\^#1{$\overline{\hbox{#1}}$}

\def\<{\langle } \def\>{\rangle }  
 
\def \({\left( } \def \){\right) }

      \let\and=\wedge
     
\def\|#1{{}_{\bigg|_{#1}}}

\def\beq{\begin{equation}} \def\eeq{\end{equation}} 
\def\bea{\begin{eqnarray}} \def\eea{\end{eqnarray}}

\def\plb#1 #2 {Phys. Lett. {\bf B#1} #2 }
\def\phr#1 #2 {Phys. Rep. {\bf  #1} #2 } 
\def\npb#1 #2 {Nucl. Phys. {\bf B#1} #2 }
\def\aph#1 #2 {Ann. Phys. {\bf #1} #2 }  
\def\jmp#1 #2 {J. Math. Phys. {\bf #1} #2 }
\def\prd#1 #2 {Phys. Rev. {\bf D#1} #2 }
\def\prl#1 #2 {Phys. Rev. Lett. {\bf #1} #2 }
\def\rmp#1 #2 {Rev. Mod. Phys.  {\bf #1} #2 }
\def\zpc#1 #2 {Z. Phys. {\bf #1C} #2 }
\def\cmp#1 #2 {Comm. Math. Phys. {\bf #1} #2 }
\def\mpl#1 #2 {Mod. Phys. Lett. {\bf A#1} #2 }
\def\ijmp#1 #2 {Int. J. Mod. Phys. {\bf A#1} #2 }
\def\jpa#1 #2 {J. Phys. {\bf A#1} #2 }

\begin{document}
{\hfill ITP--UH--14/94\vskip-9pt     \hfill hep-th/9409077}
\vskip 30mm \centerline{\hss\large \bf
    A comment on ``On Non-Abelian Duality'' by Enrique \'Alvarez, \hss}
\vskip 7mm\centerline{\hss\large \bf
    Luis \'Alvarez-Gaum\'e and  Yolanda Lozano \hss}
\begin{center} \vskip 12mm
       {\large Harald SKARKE}
\vskip 5mm
       Institut f"ur Theoretische Physik,
       Universit"at Hannover\\
       Appelstra\3e 2,
       D--30167 Hannover,
       GERMANY
\vskip 5mm
      e-mail: skarke@itp.uni-hannover.de
\vfil                        {\bf ABSTRACT}                \end{center}

The paper commented upon gives the impression that whether the gauged
version of a sigma model gives rise to the original or the dual model
depends on the choice of gauge fixing. It is demonstrated here that this is
not so.

\vfil\noindent \\ September 1994 \msk
\thispagestyle{empty} \newpage
\setcounter{page}{1} \pagestyle{plain}

In the paper [1] to which this comment refers,
sentences like ``We will show that one can understand the transformations as a
change of gauge condition'' (in the introduction) or ``The lesson to be
learned from this section is that one way
to think about duality is the equivalence of descriptions
of the same theory based on different gauges'' (at the end of section 2)
seem to give the impression that, depending on the choice
of gauge fixing in the gauged version of a sigma model, one has to end up
with either the original or the dual theory.

In the paragraphs between Eqs. (2.6) and (2.11) it is shown how the original
model can be obtained from the gauged model in the light-cone gauge and
how the dual model can be obtained in the Landau gauge. We will now show that
it is also possible to obtain the dual model in the light-cone gauge and
the original model in the Landau gauge.

In the light-cone gauge the constraint
\beq \partial_{+}\chi -\xi_i \partial_{+}\phi^i = 0   \eeq
coming from the integration over $A_{-}$ becomes, in
the coordinates adapted to the description of the isometry,
\beq \partial_{+}\chi -g_{00}\partial_{+}\theta-g_{0\a}\partial_{+}\phi^\a
         = 0.\eeq
This condition can also be used to eliminate $\theta$, and it is
easily checked that this results in the dual model with the correct
determinant.

Working in the Landau gauge, integrating out $\chi$ gives the constraint
$\partial_{+}\partial_{-}\rho=0$. On the sphere $\Delta\rho=0$ is solved
only by $\rho=const.$, so that we trivially end up with the original model.

So, whether we get the original or the dual model depends only on our
choice of which variables we decide to integrate out, and not on the
gauge we choose. It should be stressed that none of the results in the other
sections of this interesting paper are affected by these considerations.

\ni
{\bf Reference}

\ni
[1] Enrique \'Alvarez, Luis \'Alvarez-Gaum\'e and  Yolanda Lozano,
\npb 424 (1994) 155.
\end{document}